# Hyper-incursion and the Globalization of the Knowledge-Based Economy

## Loet Leydesdorff

*University of Amsterdam, Amsterdam School of Communications Research (ASCoR)*
*Kloveniersburgwal 48, 1012 CX Amsterdam, The Netherlands*
*loet@leydesdorff.net . http://www.leydesdorff.net*

**Abstract.** In biological systems, the capacity of anticipation—that is, entertaining a model of the system within the system—can be considered as naturally given. Human languages enable psychological systems to construct and exchange mental models of themselves and their environments reflexively, that is, provide meaning to the events. At the level of the social system expectations can further be codified. When these codifications are functionally differentiated—like between market mechanisms and scientific research programs—the potential asynchronicity in the update among the subsystems provides room for a *second* anticipatory mechanism at the level of the transversal information exchange among differently codified meaning-processing subsystems. Interactions between the two different anticipatory mechanisms (the transversal one and the one along the time axis in each subsystem) may lead to co-evolutions and stabilization of expectations along trajectories. The wider horizon of knowledgeable expectations can be expected to meta-stabilize and also globalize a previously stabilized configuration of expectations against the axis of time. This recursive incursion on the incursive dynamics of expectations can be modeled using hyper-incursion. The knowledge-based subdynamic at the global level which thus emerges, enables historical agents to inform the reconstruction of previous states and to co-construct future states of the social system, for example, in a techno-economic co-evolution.

## INTRODUCTION

Unlike weakly anticipatory systems, strongly anticipatory ones cannot provide predictions because these systems use their expectations for the construction of their future states.[1,2] For example, a social system reconstructs its future by providing new meaning to its past and thus potentially overwrites (parts of) its history. During the last century, for example, this reconstructive capacity has been developed particularly in terms of techno-economic coevolutions at interfaces between the natural sciences and the economy.[3] These two subsystems (science and the economy) first operate upon each other historically,[4] that is, along the axis of time, but the resulting process of continuous reconstruction can be considered at the systems level as the operation of a strongly anticipatory system. Unlike a historical development a strongly anticipatory system operates against the axis of time, that is, it operates in the present with the perspective of hindsight.

Each subsystem contains a model of the other subsystems by using its own code as an axis for the projection. For example, everything can be exchanged on the market for a price or, from a different perspective, made the subject of scientific research. The globalizing knowledge base can be considered as a hyperincursive interaction effect among these incursive mechanisms. It has been suggested that the incursive and hyper-incursive equations can be used for the modeling of weakly and strongly anticipatory systems, respectively.[1,5] This paper suggests explaining the mechanism of reconstruction and "creative destruction"[6] in the evolution of social systems in terms of these equations.

The incursive formulation of the logistic equation has been used previously to model interaction among and aggregation of (weakly) anticipatory agents.[7] These incursive models extended on epidemiological models,[8] but paid no attention to the additional processing of meaning in social systems.[9] Meaning-processing entails a reference to other possible states, i.e., a global system of reference.[10] In the case of a social system, one is able to distinguish between the stabilization of expectations (e.g., in a knowledge infrastructure) and the system's knowledge base as an order of expectations which remains pending in a virtual realm.[11]

The knowledge base operates globally on the basis of expectations generated and codified within the system itself, while a knowledge infrastructure can be considered as a locally stabilized retention mechanism. Thus, the system develops at two levels at the same time: an institutional one for the retention (historical and thus along the axis of time), and another functional one for the reflexive reorganization from the perspective of hindsight. These two subdynamics couple upon each other in the reconstructions which happen to occur. In this study, I submit an appreciation of the hyper-incursive formulation of the logistic equation as the source of the reflexive reconstruction at the level of the social system.[12,13]

## STABILIZATION, META-STABILIZATION, AND GLOBALIZATION

Stabilization of expectations can be the consequence of an interaction and potential co-evolution between two subdynamics (like "action" = "– reaction"). A co-evolution between two subdynamics ("mutual shaping") can resonate into a stable configuration. Globalization, however, requires a third mechanism (e.g., reflexive decision-making about the events) to be added to the system. For example, sequences of decisions can be "locked-in" as a systems property under the condition of network externalities.[14,15] After a lock-in the system may be considered as hyper-stabilized. However, this configuration can also be meta-stabilized by using the sign opposite to the option of hyper-stabilization (Figures 1a and 1b).

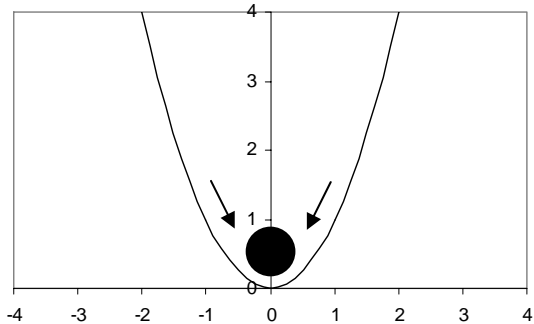

**FIGURE 1A.** Hyperbole with a minimum: stabilization and hyper-stabilization

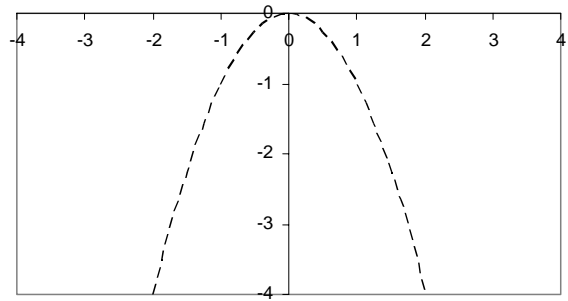

**FIGURE 1B.** Hyperbole with a maximum: meta-stabilization

In other words, equilibrium can evolve into a meta-stable system when the sign of the equation is changed. For example, the change from a market with decreasing marginal returns into one with increasing marginal returns can lead to a bifurcation at the meta-stabe vertex of the hyperbole and a consequential lock-in on either side. In the case of a lock-in, the previously meta-stable system is hyper-stabilized by using a third feedback term which reinforces the coevolution between the two subdynamics.

Which are the conditions for break-out from a lock-in?[16] A meta-stable system can experience a phase transition if a third subdynamic (with another parameter) is added at the systems level like in a triple helix model.[17] At a saddle-point (like in Figure 2a), another bifurcation—that is, other than hyper-stabilization—becomes possible. When the stabilizing and globalizing subdynamics are further differentiated—that is, provided with different parameter values—the meta-stable system may avalanche into another basin of attraction or return to its previous state (Figure 2b).

The transition towards a globalized and knowledge-based economy can be considered as such a phase transition. Because of the ongoing need of a historical realization, one expects both the globalizing and the stabilizing subdynamics to operate, and thus the system to remain in transition. (Without retention mechanisms operating, the system (e.g, the economy) would become footless and virtual instead of *globalizing*.) This "endless transition" between global options and local stabilizations can be considered as the hallmark of a knowledge-based economy or, in other words, as endogenous to the so-called triple helix of university-industry-government relations.[18]

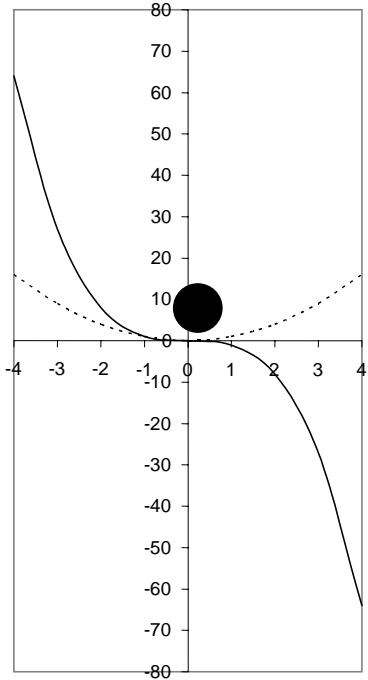 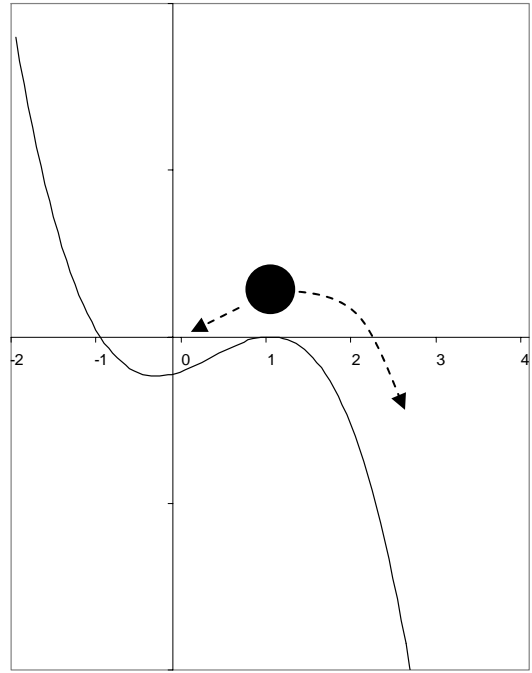

**FIGURE 2A.** The possibility of meta-stabilization by adding a third factor

**FIGURE 2B.** This system can avalanche into a new state or return to a previous state

For example, the market operates in terms of economic transactions according to its own rules. Because the economic relations are codified (e.g., using prices and currencies), the network can retain value from the exchanges. Thus, the capitalist system entertains an economic model of itself.[19] Analogously, the science system has increasingly developed its own codifications since the Scientific Revolution of the 17[th] century.[20,21] The sciences develop and differentiate rewriting their history along trajectories over time, while market clearing occurs at each moment in time. These two anticipatory mechanisms can be expected to develop along analytically orthogonal axes of the social system; they interact insofar as they are interfaced in the historical events and actions.

Historically, the interfacing of economic exchanges and scientific communications since the late 19[th] century has first stabilized a techno-economic system during the 20[th] century.[4] This coevolution had to be supported by a "technostructure"—first in the corporate sphere,[22] but then increasingly also in the public arena—because economic expectations and research perspectives tend continuously to differentiate given the differences in the codification of the respective communication systems. The integration has to be organized as another feedback. The organized interfacing of these two types of expectations provides room for hyper-incursive decision-making and the consequent coding of decision rules at the systems level. A triple helix can thus endogenously be generated.

When three subdynamics interact, their mutual information can provide the system with a negative entropy contribution, that is, a reduction of the uncertainty.[23,24] This negative entropy can be developed into an indicator of the development of a knowledge base in an economy.[25,26,27] When this indicator is measured at different moments of time a comparative static analysis can be pursued. In this study, however, I focus on the algorithmic model in order to study the complex dynamics of expectations.

## INTERACTIONS AMONG WEAKLY ANTICIPATORY SYSTEMS

How can stabilization, hyper-stabilization, meta-stabilization, and globalization of anticipation be modeled using incursive and hyper-incursive formulations of the logistic equation? An incursive formulation of this equation is:

$$x_{t+1} = ax_t(1 - x_{t+1}) \qquad (1)$$

or, after the appropriate reorganization of this formula:[28]

$$x_{t+1} = ax_t/(1+ax_t) \qquad (2)$$

Using this equation, one can generate an endogenous model of the system. Figure 3 provides an example using a cellular automaton.[9] Furthermore, one is able to generate systems entertaining different models of the observed system by changing the value of the parameter *a*. These representations can be considered as observing systems with different biases or "blind spots".[29]

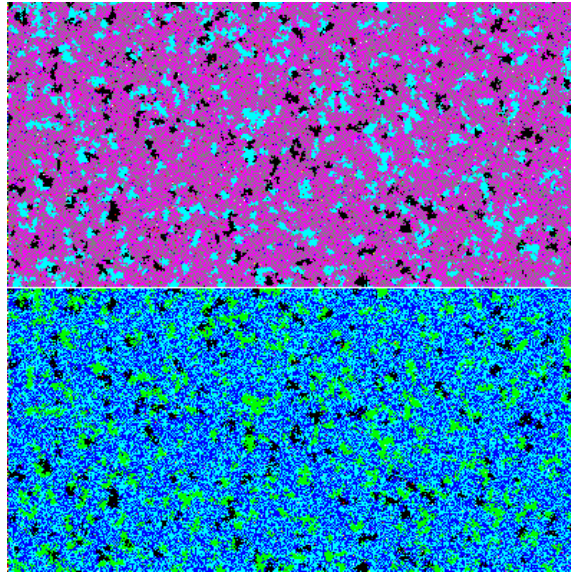

**FIGURE 3.** The top-level screen produces a representation of the bottom-level one by using an anticipatory algorithm

At the level of a social system one can expect both *aggregations of* and *interactions among* the various observations.

## Aggregation

Let us first turn to the aggregation. The sum of two observations with parameters *a* and *b*, respectively, can be formulated (on the basis of Equation 2) as follows:

$$\begin{aligned} x_{t+1} &= [ax_t/(1+ax_t)] + [bx_t/(1+bx_t)\} \\ &= [(1+bx_t)ax_t + (1+ax_t)bx_t]/[(1+ax_t)(1+bx_t)] \\ &= [2abx_t^2 + (a+b)x_t]/[abx_t^2 + (a+b)x_t + 1] \end{aligned} \qquad (3)$$

This formula is symmetrical in the parameters *a* and *b*. The equation can be considered as a hyperbole with a stable solution at the origin for $x(t) = 0$. (I will come to the second root of the equation below.) Using simulations,[9] it could be shown that an uncertainty in the representation of the modeled system is generated by the aggregation when the two modeling systems are not synchronized. The two observers use different angles (parameters *a* and *b*) and different (random) frequencies for the observation, and this can be expected to generate uncertainty in the resulting aggregate.

In the case of competing technologies, for example, one would expect that the competition among different models generates economic exchanges as another (potentially coevolving) communication system. Over time, however, this uncertainty may provide an independent source of probabilistic entropy, and thus may also act as a destabilizer of the (prevailing) equilibrium condition. The one root of the equation [$x(t) = 0$] leads to equilibrium, but the other [$x(t) = -(a+b)/2ab$] not necessarily. If the sign of the one parameter is different from the other, the orientation of the hyperbole can be reversed (Figure 1b).

Codification can be produced endogenously to the further development because of the mechanism of lock-in. When a historical destabilization is further codified at the systems level, this can lead to meta-stability. As Arthur showed,[14,15] a random walk necessarily passes the threshold of a lock-in in the case of increasing marginal returns—that is, in other words, a positive feedback. In this case, one of the parameters (*a* or *b*) is systematically set at a non-random value and a phase transition in the technological cycle thus can be induced.

In the case of a lock-in, the economic diffusion dynamics and the production rate are co-evolving.[30] However, this co-evolutionary process can again be bifurcated when the diffusion coefficient becomes considerably larger than the reaction parameter.[31,32,2:pp.182ff.] Thus, a third subdynamic—and correspondingly, a third parameter—can endogenously be generated. This parameter may induce a life-cycle in the technology by changing the sign of the feedback.[33]

In other words, the system can be expected to bifurcate and then the previous coevolutions are locked-in into a new regime, while other coevolutions are resolved during this transition. Bifurcation and lock-in thus can be considered as two sides of the same medal. A third axis (*c*) can be expected to coevolve with and bifurcate from the interaction term between the previously established subdynamics. This operation is recursive, that is, it can operate upon its own results by deconstructing and reconstructing the previously established configurations.

## Interaction

While the aggregations can be linear, interactions between two different representations of a system are non-linear. In general, interaction can be modeled as a multiplication. Multiplication of the two incursive observations leads to Equation 4 in which the coefficient *d* represents the strength of the interaction:

$$\begin{aligned} x_{t+1} &= d*[ax_t/(1+ax_t)]*[bx_t/(1+bx_t)] \\ &= d*[ax_t/(1+bx_t)]*[bx_t/(1+ax_t)]/(1+ax_t+bx_t+abx_t^2\} \\ &= [d*(ax_t+abx_t^2)(bx_t+abx_t^2)]/[1+(a+b)x_t+abx_t^2] \\ &= d*[px_t+qx_t^2+rx_t^3+sx_t^4]/[abx_t^2+(a+b)x_t+1] \end{aligned} \quad (4)$$

Given appropriate choices of the parameters, Equation 4 can resolve into a quadratic equation and therefore provide us with a minimum or a maximum (depending on the parameter values). Thus, we remain within the domain of hyperbolic functions (with stable and meta-stable solutions). An analytically independent, third subdynamic is not yet part of this model. Stabilization can be considered as the result of the dynamic extension of the interaction among two observing (that is, incursive) systems. In the case of a co-evolution among the expectations along a trajectory, the expectations can become "mutually adjusted" by recursive interactions in an otherwise less organized environment.

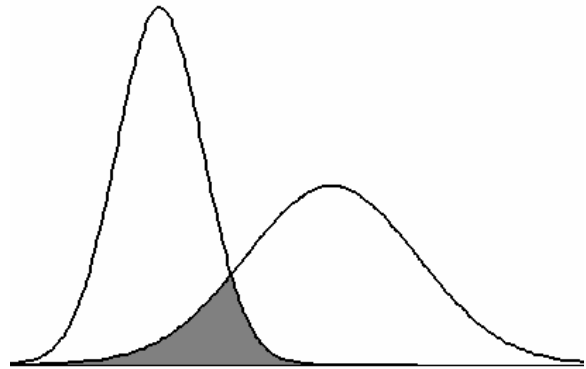

**FIGURE 4.** Two variations can stabilize a next-order variation by selecting upon each other.

Note that the results of the interaction may provide us with a richer representation than each of the interacting observations. Interaction allows the selections formalized in the right-hand term of the logistic equation (that is, the feedback term [1 - *x(t+1)*] in Equation 1) to operate upon each other in the multiplication. Two selective feedbacks operating upon each other lead to a quadratic term $(1 - 2x(t+1) + x(t+1)^2)$ that can be positive, while each selection can be considered as a negative operation. The positive term provides a second-order variation, but this variation is an order of magnitude smaller than the original variations (Figure 4). Whereas each observation contains a selection from the phenotypical complexity, some selections can be selected for stabilization. However, this assumes that the system contains two independent anticipatory mechanisms that are able to operate upon each other, for example, in a co-evolution. Let us now turn to the specification of these two anticipatory mechanisms.

## GLOBAL HYPER-INCURSION

I shall first discuss the specification of the hyper-incursive coevolution as a formal possibility and then provide it with a substantive interpretation in a next subsection.

### The Possibility of Global Hyper-Incursion

Under the condition of functional differentiation in the codes of the communication the social system can be expected to contain two analytically different mechanisms for the anticipation, namely:
1. the information exchanged can be provided with meaning by each observer or modeling system from the perspective of hindsight; these expectations can interact and stabilize a codified expectation at the level of the social system (independent of whether the system is functionally differentiated or not);
2. when the social system is functionally differentiated, the codes of the communication are no longer synchronized *ex ante*. The asynchronicities in the updates among the subsystems provide the *social system*—unlike the psychological system—with a second $\Delta t$ which can be used for anticipation.[34] The additional degree of freedom is provided in principle by the distribution of the individual observers who provide a range of reflections. However, codification can structure this dimension as another selection mechanism.

For example, the economy operates and updates with cycles very different from scientific communication systems or political elections. Synchronizations take place by providing meaning across (more or less structured) interfaces at specific moments of time, but the construction of meaning within each system takes place *along* the time axis. This historical development stands orthogonally to the transversal updates among differentiated systems at specific moments of time. The meaning processing thus provides a second layer of interactions among the reflexive agents which can interact with the information processing among them. Therefore, it can also be considered as an independent dimension or another subdynamic of the social system.

The coevolution between two analytically different mechanisms of anticipation can be expected to provide first the stability in the historically observable instantiations of a strongly anticipatory system. However, globalization operates on stabilizations as an evolutionary mechanism against the axis of time, that is, by providing a different meaning to what has been happening hitherto.[35] Whereas some selections can be selected for stabilization along the historical axis, some stabilizations can recursively be selected for globalization. This next-order selection mechanism presumes another axis within the system analytically different from the one of historical stability. While the historical stabilization is reinforced by socially interacting and aggregating incursions (Equations 3 and 4 above), hyper-incursion opens the system towards future states and thus introduces a different source of uncertainty.

Under which conditions can an organization of the expectations emerge which provides the previous coevolution between the other two subdynamics with a feedback term? As long as the parameters of the two subdynamics (stabilization and globalization) are similar, the system remains in a meta-stable configuration (Figure 2a). The new axis of globalization can emerge as independent with the codification of the uncertainty in the expectations at the interface (e.g., by developing a decision-rule). As it emerges and is stabilized (e.g., in an organization like a firm), this third subdynamic operates orthogonally to the ones on which it builds historically and from which it emerged (like at the level of a hypercycle).

Although the system remains phenotypically composed of different subdynamics, the emerging subdynamic of globalization itself remains analytical and therefore virtual in practice: it cannot appear historically without mediation. In other words, the hyperincursive system remains a hypothesis about a pending transformation or an order of expectations for the incursive systems carrying it. These historically mediating agents can be psychological systems or institutional agents because such systems can maintain a historical manifestation in the present. Because of its potential to combine reflexivity and historicity, the individual can entertain a model of a future state with reference to this system's present state as a "weak" anticipation. On the one hand, the agents need the social system for generating change because they cannot reconstruct the system themselves as weakly anticipatory systems. The social system, on the other hand, cannot be expected to exist concretely: it remains an order of expectations. Its hyper-incursive operation can be hypothesized by historical agents on the basis of observing its footprints in hitherto provisionally stabilized retention mechanisms. These hypotheses can be further informed by the theoretical models entertained and by empirical data.

Three hyper-incursive models follow from the logistic equation:

$$x_t = ax_t(1 - x_{t+1}) \tag{5}$$

$$x_t = ax_{t+1}(1 - x_t) \tag{6}$$

$$x_t = ax_{t+1}(1 - x_{t+1}) \tag{7}$$

- Equation 5 evolves into $x = (a - 1)/a$ = Constant. I submit that this evolution towards a constant through anticipation can be considered as modeling the self-reference of an (individual) identity. At the level of the social system a group can also have an identity;
- Equation 6 evolves into $x_{t+1} = (1/a) \{x_t / (1 - x_t)\}$. Since the latter term approaches –1 as its limit value and the former term is a constant $(1/a)$, this representation can alternate between itself and its mirror image. This subdynamic thus formalizes the reflexive operation. Reflection is relevant both in the case of an identity (e.g., as the organization of an individual) and in the case of a social system that contains in addition to the aggregation also the interaction terms;
- Equation 7 is only possible on the basis of interaction between and among psychological systems because this system $(x_t)$ has lost any reference to the present state. It only refers to its own future state and its own being selected in this future state, but this system cannot exist otherwise than in terms of expectations.

## Intersubjective Anticipation

The intersubjective intentionality modeled by Equation 7 has been theorized in sociology as the "double contingency" between *Ego* and *Alter*.[36] Using this mechanism for the anticipation, the *Ego* in the present $(x_t)$ no longer has a reference to itself, but only to itself in a future state $(x_{t+1})$, that is, as an *Alter Ego*. The orientation of providing meaning with reference to other possible meanings thus constitutes the social world as a system different from psychological ones.[37] Unlike social systems, psychological systems contain a reference to themselves as an identity in the present (Equation 5). However, the social system of expectations remains structurally coupled to psychological systems because it can otherwise not be manifest historically.[12] The social system is only perceptible after a reflexive turn.[10]

The coupling between the social system and agents carrying it is provided by the reflexivity formalized in Equation 6. Because of this reflexivity the social system can be entertained by individuals as a notional order. *Vice versa*, the individuals or other agents are reflected in the social system as addresses or nodes. While both systems can contain the reflexive operation expressed in Equation 6, a coevolution ("mutual adjustment") between the social and the psychological systems can also be expected. But this coevolution can be interrupted when expectations are further codified on either side beyond control for the other side. For example, an eigendynamics at the level of the social system may lead to "alienation." while the recursive application of the reflexive operation provides discretionary room for private thinking and tacit knowledge on the side of psychological systems. In summary, both systems can also apply their reflexivity on themselves and thus develop hyper-reflexivity. Under pre-modern conditions, however, the social system was not yet sufficiently set free to use this degree of freedom as a differentiation because of the need of central control. This has changed during the past centuries.[13]

Let me emphasize that all next-order systems contain the lower-order subdynamics on which they rest. For example, the social system also contains its manifestations (e.g., social institutions which develop historically with the axis of time like in the epidemiological models which were discussed in a previous section). The reflexive layers of psychological and social systems are evolutionary achievements which are added as subdynamics to the system. The reflexive layer first refers to a selection by the system as an identity (Equation 5), but then also as an expectation of an identity (Equation 7). This *Ego* is reaffirmed by the selection by *Alter*—that is, $(1 - x_{t+1})$—or, in other words, by providing a mutual (or "double") contingency. Equation 6 reaffirms the identity as the reflexive operator of the uncertainty like in Descartes's *Cogito ergo sum*. The expectation of an identity in the future (Equation 7) presumes an *Alter Ego* (or the expectation of a future *Ego*) to take part in an intersubjective intentionality.[10] This can also be discussed as "trust" or "social capital."

Although the systems are constructed bottom-up, control operates top-down. Thus, once the social system of inter-human communication is in place as an order of intersubjective intentionality, the reflection of future states at the psychological level is controlled from this next-order level.[38] Unlike a social system, an individual can reflect on the future given one's current state. However, one needs others to reconstruct the system because psychological systems are only weakly anticipatory. One's historical continuity, however, provides room for action on the basis of predictions and discretionary solutions in the reflexive domain (like private meaning and tacit knowledge).

The solutions which were found hitherto can sometimes be retained and stabilized at the biological level of the individual's body (which is structurally coupled to the mind at the lower end). Analogously, the social system is "embodied" in the social institutions. Institutions provide the system with physical continuity. Like individuals, institutions and organized collectives have room for agency and tacit knowledge. The hyperreflexive uncertainty thus can be black-boxed because of the continuity and the reflexivity contained at each systems level. However, the specific subdynamic without such historical continuity in its manifestation has to remain a knowledge-based order of expectations.

## The Hyper-Incursive Operation

Equation 7 enables us to formalize the hypothesis about the global dimension of the knowledge-based order of social systems as follows:[1,39]

$$x_t = ax_{t+1}(1 - x_{t+1})$$
$$x_t = ax_{t+1} - ax_{t+1}^2$$
$$ax_{t+1}^2 - ax_{t+1} + x_t = 0 \qquad (8)$$
$$x_{t+1}^2 - x_{t+1} + x_t/a = 0$$

For $a = 4$, $x_{t+1}$ can mathematically be defined as a function of $x_t$ as follows:

$$x_{t+1} = 1/2 \pm 1/2 \sqrt{(1 - x_t)} \qquad (9)$$

Note that this formula is no longer rooted in the historical arrow of time.[40] This system is unpredictable in the sense that it is not possible to compute its development by knowing its historical conditions. As the system can only take one value at each time step, another (decision) mechanism must be made available to resolve this problem in the historical progression. Since the mechanism is no longer rooted in history, it has to be knowledge-based. For example, Dubois proposed to add the following decision function u(t) for making a choice at each time step:[39:pp.114f.]

$$u_t = 2d_t - 1 \qquad (10)$$

where $u = +1$ results from the decision $d = 1$ (true) and $u = -1$ from the decision $d = 0$ (false). As Dubois noted, the decisions $d_t$ do not influence the dynamics of $x_t$, but only guide the system which itself creates the potential futures.

In the social system, this guidance is provided by the *historical organization* of the system. Decisions can stabilize an organization over time, and decision-rules can be codified given an organizational context. Both the stabilizing organization (for example, in institutions) and the globalizing self-organization of the fluxes of communications in different directions belong to the social system as subdynamics. The subdynamics, however, should not be reified and then confused with the system. The self-organization of the fluxes in the globalizing system of communications is constrained and guided by the historical organization of the stabilizing subdynamic, but one can expect that this next-order level will increasingly exert control. Since the rewrite cannot realize itself historically, that is, without mediation, the two subdynamics can be expected to operate upon each other as different selection mechanisms. Giddens called this oscillation "the duality of social structure."[11] In evolutionary terminology, the various subdynamics operate upon each other as mutual selection mechanisms.[13]

Historical organization can be considered as one among a variety of subdynamics of the self-organization of communication in a knowledge-based system. Organization can provide the retention mechanism for solving puzzles among the fluxes at the interfaces among (functionally) differentiated subsystems. Organization thus can also be considered as a function of the system of social coordination among others (such as economic wealth generation). The code of the communication in this organizational subsystem was theorized by Luhmann as decision-making.[41,42] Recursive decision-making by communications among reflexive agents enables the system to stabilize an expectation along a trajectory in terms of boundaries within an otherwise unorganized or differently organized environment.

The social system develops at the next-order level of expectations. The order of expectations can be reflected by individuals and organizations, but this control level is one level down from expectations which are codified at the intersubjective level (e.g., in a discourse). The global level drives the knowledge-based economy as an order of expectations, but it fails to achieve this driving because it cannot succeed in the materialization without the reduction of the complexity by (individual and institutional) agents. Thus, the strongly anticipatory system remains in transition between these two levels. The levels should not be reified, but be considered as subdynamics that operate upon one another like stabilization and globalization.

## CONCLUSION

The interaction of two anticipatory mechanisms allows for coevolution and stabilization, but additionally for meta-stabilization and globalization using a hyper-incursive routine. The hyper-incursion can develop into a third axis of codification if decision-rules coevolve (e.g., into governance) among the subsystems which are organized as anticipation at the level of the social system. A triple helix can thus endogenously be generated. Historically, the interfacing of economic and scientific communications since the late 19th century has first stabilized a techno-economic system during the 20th century. This coevolution had to be supported by a "technostructure" because economic expectations and research perspectives tend to stand orthogonally. The organized interfacing of these two types of expectations has provided room for hyper-incursion and the consequent development of incursive decision rules at the systems level.[39]

Decisions first induce a local trajectory in a global space of other possibilities. When this trajectory is reflexively recognized as one among other possible trajectories, decisions can be made the subject of discursive analysis and consequential codification. Three subdynamics thus interact: (1) economic wealth generation, (2) systematic novelty production (R&D), and (3) structuration of the decision-making at the interfaces. The knowledge-based subdyanamics which emerges, reconstructs previous states and co-constructs future ones from a global perspective. The knowledge-based options are traded-off against the historical retention of wealth in the economy by making decisions in the present, but in an increasingly anticipatory mode.